# BRIDGE THEORY: OLTRE LA FRONTIERA QUANTISTICA

**Massimo Auci**[*]

*OdisseoSpace*
*Via Battistotti Sassi 13, 20133 Milano*

## 1. – Introduzione:

Proposto da Louis de Broglie nel 1924, il dualismo onda - corpuscolo aveva lo scopo di rendere applicabile anche alle particelle materiali l'ipotesi einsteiniana sul carattere ondulatorio e corpuscolare della radiazione elettromagnetica. Nelle sue linee essenziali, il dualismo assegna a luce e particelle una simultanea natura ondulatoria e corpuscolare, giustificando perciò l'uso simultaneo di teorie strutturalmente differenti come la teoria della Relatività e la Meccanica Ondulatoria.

Sin dal 1927 il dualismo venne sottoposto a severi test sperimentali, utilizzando sia fasci di luce che di particelle. I risultati hanno sempre confermato due diverse realtà fisiche che possiamo così sintetizzare: (a) l'effetto fotoelettrico studiato da Einstein (1905) e l'effetto Compton (1923), sono compatibili con una natura corpuscolare della luce e della materia, dimostrando che la luce è costituita da quanti elementari, i "fotoni"; (b) due fasci secondari prodotti nella suddivisione di un unico fascio primario monocromatico, se nuovamente sovrapposti dopo aver fatto percorrere a ciascun fascio un percorso lievemente differente, compongono su uno schermo, indipendentemente dalla loro originaria natura materiale o elettromagnetica e dall'intensità del fascio primario, una figura di interferenza costituita da una successione di frange chiare e scure. Ciascuna figura appare costituita da una sovrapposizione di singoli impatti puntiformi.

Mentre i risultati (a) contribuiscono alla conferma sperimentale del principio di quantizzazione dell'energia introdotto da Max Planck nel 1900, confermando per la luce una natura corpuscolare e non ondulatoria come previsto dalla teoria elettromagnetica di Maxwell, i risultati (b) evidenziano sia per la luce che per le particelle materiali una simultanea natura ondulatoria e corpuscolare.

Sino ad oggi tutti gli esperimenti di interferenza effettuati hanno sempre confermato il dualismo onda-corpuscolo, permettendoci di accettare oltre ogni possibile e ragionevole dubbio quantizzazione e dualismo come principi fondanti del pensiero quantistico e di ogni altro suo sviluppo contemporaneo.

Nel 1983, cinquantanove anni dopo la pubblicazione del suo lavoro sul dualismo onda-corpuscolo, Louis de Broglie in un articolo pubblicato su "Histoire Générale des Sciences" [1], giustifica le scelte fatte inquadrandole in un contesto fisico e storico d'epoca: *"lo scopo essenziale (…) era arrivare ad una teoria sintetica delle onde e della materia in cui i corpuscoli apparissero come un comportamento particolare di una struttura ondulatoria controllati dalla sua propagazione (...) alcuni indizi suggerivano proprio questa via: la teoria di Hamilton-Jacobi, sviluppata (…) nel quadro della meccanica analitica classica sembrava indicare una stretta parentela fra i moti dei punti materiali e la propagazione di un'onda; l'intervento dei numeri interi nelle formule di quantizzazione della vecchia teoria dei quanti facevano pensare che fenomeni di interferenza o di risonanza intervenissero nella stabilità dei moti degli elettroni atomici ecc. Fu ispirandomi a queste particolarità che potei gettare le prime basi della meccanica ondulatoria, e ottenere con l'aiuto di concetti relativistici, le relazioni che legano l'energia e la quantità di moto di un corpuscolo alla frequenza e lunghezza d'onda dell'onda che le ipotesi della meccanica ondulatoria portavano ad associargli…"*. Chiaramente, in un'epoca di transizione posta a cavallo tra il determinismo relativistico e l'indeterminismo quantistico, Louis de Broglie non avrebbe per nessun motivo potuto fare a meno di prendere in considerazione le teorie più attuali e brillanti della sua epoca, anche se entrambe ignorano l'aspetto della compatibilità concettuale e formale, soprattutto in termini di località del corpuscolo materiale e di non località dell'onda che lo descrive.

Solo un anno dopo l'articolo di de Broglie, in occasione di un convegno di Meccanica Quantistica tenutosi ad Amalfi, Jhon Bell a cui era stato affidato il compito delle conclusioni disse: *"…siamo in presenza di una evidente profonda incompatibilità tra i due pilastri su cui si basa la scienza contemporanea, (la Teoria della Relatività e la Meccanica Quantistica). Attendo con impazienza le tavole rotonde in cui si lasceranno da parte gli sconvolgenti dettagli tecnici degli ultimi sviluppi, per riflettere su questa strana situazione. Forse una vera sintesi tra la Meccanica Quantistica e la Teoria della Relatività non ha bisogno solo di progresso tecnico ma di un radicale rinnovamento concettuale."*

---

[*] e-mail: massimo.auci@odisseospace.it

Rinnovamento concettuale che ha radici proprio nel teorema di Bell che dimostra come due particelle identiche sono correlate indipendentemente dalla distanza che le separa. Supponiamo perciò di avere un sistema con due particelle di spin opposto, uno up e uno down. Lo spin delle particelle è indipendente dalla distanza. Utilizzando un campo magnetico per modificare l'orientamento dello spin di una delle due particelle, anche lo spin dell'altra si modificherà di conseguenza, ma nella direzione opposta.

Questi risultati sono stati confermati da due esperimenti storici, il primo eseguito nel 1972 da John Clauser e Stuart Freeman negli Stati Uniti e il secondo da A. Aspect, P. Grangier e C. Roger al CERN nel 1981. Per quanto possa apparire inconcepibile, esiste una forma di comunicazione istantanea tra le due particelle, tale che modificando lo spin di una, anche lo spin dell'altra cambia. Il concetto di cambiamento istantaneo implica però l'esistenza di una qualche forma di comunicazione a velocità infinita tra le particelle, fenomenologia attualmente non in accordo con il principio fondamentale della Relatività che prevede che la velocità della luce sia un limite invalicabile, ma alla base degli esperimenti di teletrasporto nei quali mediante il processo di intrappolamento (entanglement) è possibile replicare fotoni e atomi distruggendo gli originali e ricreandone le caratteristiche a distanza. Un fenomeno che Albert Einsten si divertiva a deridere definendolo "*una fantomatica azione a distanza*".

La rottura concettuale e fenomenologica tra fisica relativistica e quantistica, sembra essere perciò un elemento di disturbo nel quadro della fisica contemporanea. Proprio dietro il dualismo onda-corpuscolo potrebbe infatti celarsi la frontiera tra il mondo microscopico, obbediente alle leggi della meccanica quantistica e il mondo macroscopico governato dalle leggi della meccanica, dell'elettromagnetismo e dalla gravità. La sfida ora è trovare e abbattere questa frontiera.

## 2. – La frontiera

Nella teoria elettromagnetica classica il dipolo rappresenta sia il primo approccio allo studio di un sistema elementare di cariche in interazione, sia la più semplice delle sorgenti.

L'analisi di un dipolo può essere affrontata a vari livelli di complessità. Solitamente si assume un dipolo formato da una coppia di cariche in interazione a distanza $R$, senza porsi il problema delle condizioni dinamiche delle particelle cariche che originano la sorgente.

Per analizzare l'emissione di un dipolo consideriamo come variabili fondamentali la sua estensione spaziale, determinata dalla distanza di interazione $R$ e la distanza $r$ di un osservatore dal centro della sorgente. Al crescere di $r$ si possono individuare tre distinte regioni: una vicina (A), delimitata spazialmente dalla corona sferica di raggio interno $R/2$ ed esterno $\lambda$, che racchiude la zona compresa tra le cariche e il primo fronte d'onda sferico della sorgente; una di induzione (B), per distanze dell'ordine della lunghezza d'onda; una di radiazione (C), per distanze più grandi della lunghezza d'onda.

In base al moto delle cariche, un osservatore nella regione di radiazione percepisce l'emissione di dipolo come uno o più impulsi elettromagnetici con una struttura in frequenza ben definita. Per poter però assumere l'esistenza di un osservatore macroscopico in grado di percepire un segnale elettromagnetico in tutte e tre le regioni (A-B-C), occorre pensare ad una sorgente in grado di contenere l'osservatore. Questo è possibile solo se la sorgente emette con una lunghezza d'onda $\lambda$ molto più grande dell'estensione spaziale $R$ del dipolo. In questo caso la lunghezza d'onda è molto più grande della minima distanza di interazione tra le cariche: $Z = R/\lambda \ll 1$ e il volume occupato dal dipolo diventa trascurabile rispetto a quello occupato dal suo fronte d'onda. Sotto queste condizioni il dipolo può essere pensato come una sorgente di estensione puntiforme, formata da una coppia di cariche dotata di moto relativo sufficientemente lento da generare un'onda di lunghezza $\lambda$ tanto grande da poter contenere l'osservatore macroscopico.

L'onda emessa dal dipolo non è perfettamente sferica, ha la forma di un ellissoide con un campo elettromagnetico effettivo a simmetria cilindrica e asse coincidente con l'asse del dipolo, ma per $Z \ll 1$ può essere approssimata a simmetria sferica, in quanto il vettore di Poynting associato al campo elettromagnetico della sorgente è a meno di una componente trasversale trascurabile, normale ad un fronte d'onda sferico.

All'epoca dei miei studi universitari mi incuriosì il fatto che gli atomi di idrogeno possono essere considerati dei sistemi dipolari con un momento determinato dalla distanza fra nucleo ed elettrone. Negli atomi neutri eccitati, l'elettrone si allontana dal nucleo tanto di più quanto è maggiore l'energia assorbita. Assumendo arbitrariamente una descrizione del modello atomico puramente classica, si può affermare che il momento di dipolo di un atomo di idrogeno eccitato è maggiore di quello di un atomo nello stato fondamentale. Quando l'atomo si diseccita, la distanza fra le cariche torna al valore fondamentale, perciò durante la transizione tra i due stati, l'atomo deve emettere l'energia in eccesso tramite un singolo impulso elettromagnetico. L'atomo dovrebbe perciò essere una sorgente di dipolo ed emettere un'onda di energia proporzionale al quadrato dell'ampiezza del campo elettrico, invece l'emissione avviene in ben altro modo, l'energia è pari a quella di un fotone di energia $E = hc/\lambda$.

Se in un dipolo macroscopico qualunque si impone una variazione nella distanza d'interazione analoga a quella prodotta nell'atomo di idrogeno durante un salto elettronico, l'emissione è esattamente uguale a quella prevista dalla teoria elettromagnetica classica. Perché sorgenti simili che emettono su scala differente hanno comportamenti tanto diversi?

La risposta potrebbe essere quella consueta se consideriamo che l'elettromagnetismo è in grado di descrivere solo i campi prodotti dal moto collettivo di un grande numero di cariche elettriche ma non l'interazione microscopica, per esempio tra nucleo ed elettrone o tra coppie di particelle cariche; potrebbe però essere molto meno scontata se affrontiamo il problema da un differente punto di vista.

Per esaminare il problema nel modo più realistico possibile, consideriamo un dipolo che emetta con lunghezze d'onda dell'ordine della distanza d'interazione delle cariche che lo formano. In questo caso le condizioni per considerare la sorgente come se fosse a simmetria sferica ($Z \ll 1$) non sono verificate perché la distanza $R$ e la lunghezza d'onda $\lambda$ sono confrontabili $Z \approx 1$. Il vettore di Poynting del campo, oltre alla componente radiale che descrive il flusso energetico nella direzione dell'osservatore, possiede una componente trasversale non irrilevante che localizza energia nell'intorno della sorgente. Il contributo energetico non è trascurabile se non accettando di perdere informazioni sia sulla effettiva struttura del campo elettromagnetico locale, sia sulle caratteristiche dinamiche della sorgente elettromagnetica prodotta. Perciò in una sorgente reale di dipolo la componente trasversale del vettore di Poynting ha buone chances di essere la frontiera che cercavamo [2-3].

## 3. – La sorgente reale: il modello

Per ottenere previsioni attendibili sia da un punto di vista fenomenologico che quantitativo, occorre pensare ad un modello di sorgente che sia simile ad un atomo di idrogeno ma di facile gestione teorico-formale. Per evitare poi la dipendenza dell'energia emessa dalle condizioni dinamiche iniziali delle particelle che la formano e l'uso a priori di principi relativistici estranei alla teoria elettromagnetica, si è considerato un dipolo formato da una coppia di particelle prive di massa a riposo: solo una coppia di cariche. Inoltre, per evitare di imporre inavvertitamente condizioni derivanti da eventuali mixing sui valori delle costanti elettromagnetiche fondamentali, si è preferito non fissare il valore di carica elementare e ipotizzare velocità relative tra le cariche sempre inferiori alla velocità della luce.

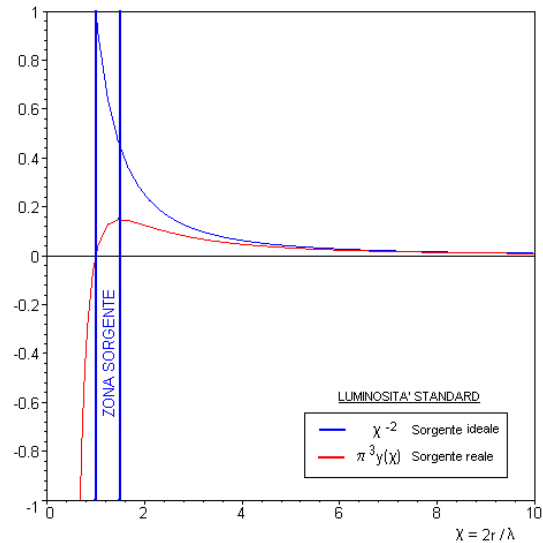

Figura 1

Il modello che ne deriva, rappresenta un'interazione "libera" tra cariche di segno opposto in movimento. Perciò a prescindere dalle masse di riposo delle particelle che non sono state assegnate, la sorgente si adatta a descrivere numerose situazioni fisiche quali le interazioni tra protone ed elettrone, ioni e coppie di particelle, offrendo il vantaggio di non imporre alcuna condizione sulle forze che generano il moto relativo tra le cariche.

L'analisi delle variabili in gioco, evidenzia che l'unico parametro dal quale dipende realmente il comportamento dinamico della sorgente è la distanza d'interazione $R$ tra le cariche. Per evitare ogni possibile arbitrarietà, occorre perciò valutare l'intervallo dei valori di distanza d'interazione entro i quali è contenuta la sorgente.

Il flusso di energia (fig. 1) in funzione della distanza dell'osservatore dal centro della sorgente, è anomalo rispetto a quello di una sorgente ideale. L'analisi dettagliata della luminosità in funzione della profondità d'osservazione $\chi = 2r/\lambda$, correlata con la variabile $Z$, mostra che durante l'avvicinamento delle cariche la luminosità cresce da zero sino ad un massimo: luminosità zero in corrispondenza di χ = 1 quando $Z = 3/2$; luminosità massima per χ = 3/2 quando $Z = 1$. Per valori $Z > 3/2$, corrispondenti a profondità di osservazione χ < 1, il dipolo non ha le caratteristiche di una sorgente, in quanto assorbe energia anziché emetterla e la luminosità è negativa (fig. 1).

Limitatamente alla fase di avvicinamento delle cariche (fig.2), se le distanze d'interazione sono tali da avere $1 \leq Z \leq 3/2$, possiamo riconoscere una fase di produzione e localizzazione dell'energia detta anche "alfa"; il valore massimo dell'energia localizzata è determinato dal valore della minima distanza d'interazione tra le cariche: $R = \lambda$. Possiamo dire che la sorgente per mezzo della sua lunghezza d'onda mantiene la memoria delle condizioni dinamiche iniziali delle particelle che l'hanno prodotta[1].

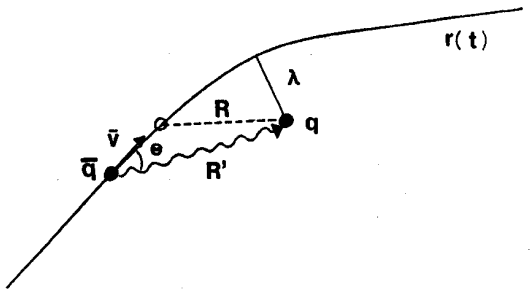

Figura 2

Superata la minima distanza d'interazione, le cariche iniziano ad allontanarsi: è la fine della fase *alfa* e l'inizio della fase "omega" che porta a distanza infinita alla totale distruzione della zona sorgente. Nella fase *omega* la luminosità riducendosi si accosta a quella di una sorgente a simmetria sferica (fig. 1).

La fase *alfa* delimita la zona sorgente ad una corona sferica compresa tra le superfici di raggio $3\lambda/4$ e $\lambda/2$: quando due cariche in moto relativo avvicinandosi raggiungono la reciproca distanza $3\lambda/2$, il dipolo emette radialmente raggiungendo la massima luminosità quando le cariche si trovano alla minima distanza d'interazione $\lambda$. La durata temporale della fase *alfa* è calcolabile solo a posteriori, cioè dopo che le particelle hanno raggiunto la distanza minima d'interazione: $\tau = \lambda/2c$. Pertanto, il tempo di "collisione" durante il quale si ha produzione di una sorgente non è infinito ma limitato. Considerando l'evoluzione della sorgente nel caso di inversione temporale del moto delle cariche, si assume che la collisione abbia complessivamente una durata doppia e si estenda simmetricamente sulla fase *omega*: $t_{col} = \lambda/c$. Il tempo di collisione oltre a coincidere con il periodo della sorgente, è il tempo necessario per la formazione del fronte d'onda sul quale è realizzato il bilancio energetico tra energie emessa e non ancora emessa [3-4].

### 4. – L'origine della quantizzazione

Il modello ad interazione libera permette di considerare la formazione di una sorgente indipendente per ogni coppia di cariche in interazione. Per esempio se un elettrone collide con un nucleo di elio, si formeranno due sorgenti, ciascuna delle quali contribuirà a localizzare energia e quantità di moto. Infatti, mentre la componente radiale del vettore di Poynting di ogni sorgente è associata al flusso di energia nella direzione di un osservatore, la trasversale è associata ad un'onda in rotazione intorno al centro virtuale della sorgente [2-3-4-5].

Funzione di Struttura

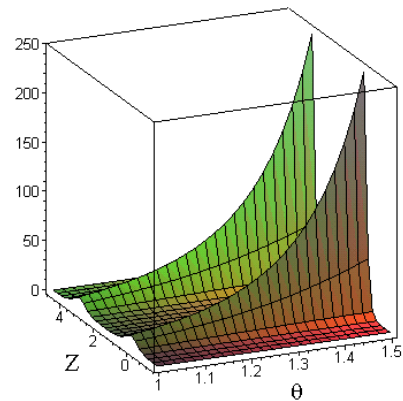

Figura 3

Nel 1980, la curiosità di giungere rapidamente ad un risultato concreto che desse informazioni sulla reale natura fisica della costante di Planck era tanta, che provai a tentativi a valutare l'energia localizzata nella zona sorgente dall'onda trasversale.

Per impazienza non mi preoccupai troppo né del valore da attribuire alla variabile $Z$, né dell'accordo formale con la Meccanica Quantistica. Mi dedicai solo al calcolo dell'energia localizzata nella zona sorgente. Fu un bene perché in questo modo non condizionai i risultati a ciò che mi sarebbe piaciuto ottenere. L'energia risultava calcolabile a partire dalla conoscenza della funzione di struttura del campo della sorgente (fig. 3) e dal tempo di collisione.

---
[1] Si veda paragrafo 6.

Formalmente l'energia localizzata risultava proporzionale ad una azione e alla frequenza effettiva di emissione della sorgente $\nu = 1/t_{col}$, in accordo con quanto previsto per l'energia di un fotone:

$$E_{loc}(Z,\nu) = \left[\frac{4\pi}{3}\frac{q^2}{c}\int_0^\pi \Theta_t(Z,\theta)\,d\theta\right]\nu \quad (1)$$

Assumendo come carica elementare quella dell'elettrone, l'azione seppur variabile in funzione di Z, per valori interni all'intervallo della zona sorgente risultò a meno di un fattore tre non allontanarsi troppo dal valore atteso per la costante di Planck. Questi risultati, anche se lontani dall'essere quantitativamente esatti, erano un primo successo.

L'esplorazione del comportamento della sorgente nell'intervallo dei possibili valori del parametro d'interazione, sia internamente che esternamente alla zona sorgente, mise in evidenza alcune peculiarità della funzione di struttura. Nel caso in cui Z viene fatto tendere a zero ($Z << 1$) o crescere per valori $Z > 2$, la sorgente si comporta in modo curioso. Nel primo caso le cariche sono talmente vicine da annullare la componente trasversale del vettore di Poynting: il dipolo si comporta come una sorgente a simmetria sferica. In queste condizioni scompare ogni forma di localizzazione e tutta l'energia prodotta dalla sorgente viene emessa radialmente dal vettore di Poynting. Il secondo caso invece non corrisponde ad una situazione accettabile, ma è comunque interessante per eventuali implicazioni che l'effetto potrebbe avere se venisse sperimentalmente evidenziato. Infatti, essendo

$$Z = \frac{R(t)}{\lambda} \cong \left(1 + \frac{v}{c}\cos\theta\right) \quad (2)$$

per $Z = 2$ la funzione di struttura è singolare perché la velocità della particella è uguale a quella della luce nel mezzo, mentre per $Z > 2$ che corrisponde a interazioni tra particelle in moto relativo a velocità superluminale, è prevista la formazione di una zona sorgente analoga a quella per valori subluminali. Questo aspetto della teoria non è però stato ancora approfondito.

Un semplice esempio di accordo teorico con realtà fisica, si ha considerando l'equazione (2) per un mezzo materiale con indice di rifrazione n. Il valore limite $Z = 2$ comporta per una particella in moto a velocità superiore a quella della luce nel mezzo, ma inferiore a quella della luce nel vuoto, collisioni con la materia che portano alla formazione di sorgenti con un angolo di emissione

$$\theta = ar\cos\frac{1}{n\beta}. \quad (3)$$

che soddisfa la condizione Cherencov.

L'energia (1) associata alla funzione di struttura all'interno della zona sorgente e il suo accordo formale con quanto previsto per l'energia di un fotone, permettono di giustificare il principio di quantizzazione con la formazione di sorgenti purché entro precisi intervalli della variabile d'interazione Z, al di fuori dei quali è la fisica del mondo macroscopico per la quale la quantizzazione è un effetto trascurabile, a prevalere. Notando che per la simmetria assiale del campo di una sorgente e per la casualità della posizione dell'asse di dipolo rispetto ad un qualunque osservatore, il vettore di Poynting che maggiormente caratterizza la sorgente è quello relativo all'angolo medio d'emissione <θ> = π/4, si è potuta valutare la distanza quadratica media tra le cariche con la quale determinare la costante di struttura del campo elettromagnetico al variare di Z (fig. 4).

Nel 1984 l'arbitrarietà del problema era stata eliminata e la strada verso il calcolo dell'effettiva costante di Planck e la comprensione del profondo significato della quantizzazione era ormai sgombra da difficoltà.

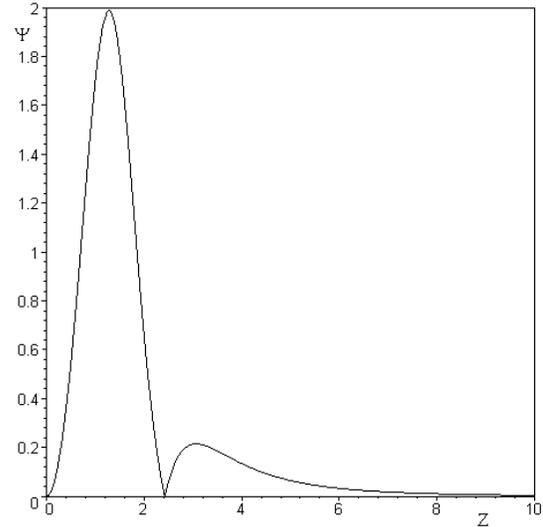

Figura 4

Il calcolo numerico della distanza quadratica media di interazione, corretto sulla base dei risultati sperimentali [4-5], mette in evidenza un fenomeno curioso. Indipendentemente dal valore effettivo della distanza minima d'interazione e dalla velocità relativa tra le cariche, ad un dipolo è sempre associata una sorgente caratterizzata da un rapporto costante tra distanza quadratica media e minima distanza d'interazione:

<Z> = 1.275556687 ± 1.1 $10^{-8}$.

Questo risultato assicura che una qualunque coppia di cariche genera sempre una sorgente dipolare, ma assicura anche che ogni carica elettrica può formare una sorgente con ogni altra anticarica con la quale si produca almeno una fase *alfa* indipendentemente dalla distanza effettiva a cui le cariche si trovano. Questo aspetto del tutto nuovo, potrebbe essere la chiave per giustificare i risultati derivanti dalla disuguaglianza di Bell, in quanto la formazione della sorgente costituisce un legame causale assoluto e permanente che lega fra loro le caratteristiche dinamiche delle particelle che l'hanno originata (vedi par. 8).

### 5. - Le costanti di Planck e di struttura fine

Lo sviluppo formale del modello ha permesso di ottenere risultanti interessanti e sicuramente nuovi nell'ambito della comprensione del legame tra fenomeni elettromagnetici classici e fenomeni quantistici. Formalmente, il calcolo dell'energia e della quantità di moto associate alla componente trasversale del vettore di Poynting nella zona sorgente, produce per energia e quantità di moto risultati pienamente in accordo con quanto previsto per un fotone in Meccanica Quantistica:

$$E = 2\pi \hbar c/\lambda \; ; \; P = 2\pi \hbar/\lambda,$$

con $\hbar = h'/2\pi$ costante di Planck teorica calcolata in funzione del parametro d'interazione $<Z>$.

I risultati ottenuti non si fermano però ad una semplice uguaglianza formale. Il valore della costante d'azione è numericamente ottenuto a meno del rapporto $q^2/c$, calcolabile in base al valore di carica elementare adottato nel modello.

A partire dall'energia (1), dalla conoscenza della distanza quadratica media di libera interazione che caratterizza un dipolo di lunghezza d'onda λ e dall'incertezza sulla distribuzione angolare delle cariche, la funzione di struttura comprensiva del termine di Lorentz, relativo all'interazione diretta fra le cariche, assume la forma:

$$A = \left( \frac{4\pi}{3} \int_0^\pi \Theta_t(\theta, <Z>) d\theta + \frac{1}{4\pi <Z>^2} \right)^{-1}$$

con valore numerico adimensionale

$$A = 1/(137.035989 \pm 6 \cdot 10^{-6})$$

in pieno accordo con i valori sperimentali della costante di struttura fine α.

Se si assume come carica elementare della coppia quella dell'elettrone, la costante d'azione è conseguentemente fissata dalla relazione

$$\hbar = A^{-1} e^2 /c$$

che la lega alla costante di struttura fine di Sommerfeld e la fa corrispondere numericamente al valore sperimentale della costante di Planck.

A differenza di quanto è stato sino ad ora, la costante di struttura fine in questo contesto è ottenibile per via teorica a prescindere dal valore delle costanti elettromagnetiche che solitamente la definiscono. Questa circostanza sembra privare la costante universale di Planck del ruolo privilegiato di costante fondamentale, che a pieno diritto passa alla costante di struttura fine, l'unica ad avere il privilegio di poter essere fissata dalla stessa teoria che la prevede.

### 6. - "Bridge Theory", un ponte tra fisica classica e meccanica quantistica

La corrispondenza formale e quantitativa tra energie e impulsi di una sorgente dipolare e di un fotone di pari lunghezza d'onda, permette di dare una spiegazione a molti aspetti della fenomenologia quantistica. Due cariche in moto relativo posseggono intrinsecamente un'energia cinetica e una quantità di moto determinabili univocamente a partire dalla conoscenza delle loro masse e delle loro velocità rispetto ad un osservatore. Più grande è la componente della quantità di moto della particella incidente lungo l'asse d'interazione, minore è la distanza minima λ raggiunta dalle cariche durante la fase *alfa* e maggiore è l'impulso e l'energia localizzata nella zona sorgente. Il dipolo ha perciò la proprietà di memorizzare e sintetizzare proprio nel valore della distanza minima d'interazione, quindi nella lunghezza d'onda, le condizioni dinamiche che determinano univocamente l'energia e l'impulso di un fotone a prescindere dai valori effettivi delle masse delle particelle che lo generano.

La massa di una particella è misurabile solo mediante un'interazione che metta in evidenza le sue caratteristiche inerziali, perciò possiamo parlare di massa solo se siamo in grado di interagire con essa. Per poter considerare la massa delle particelle di un dipolo, occorre allora che queste siano soggette ad una reciproca interazione, cosa che durante la formazione di una sorgente o durante il processo di cattura elettronica da parte di un nucleo [9], accade sempre. Nonostante ciò la massa delle particelle non compare nel modello elettromagnetico di sorgente, sembrando apparendo inessenziale per la valutazione dell'energia e della quantità di moto.

A causa del moto relativo delle cariche, la sorgente in formazione localizza un'energia e una quantità di moto descritte da un fotone di lunghezza d'onda pari alla minima distanza d'interazione raggiunta. Se al limite di un moto relativo con velocità tendente a zero le cariche fossero reciprocamente immobili, l'interazione sarebbe statica, la durata del tempo di collisione infinita e l'energia e la quantità di moto localizzate nella zona sorgente sarebbero nulle. In ogni altra situazione, la coppia di cariche in moto possiede invece un'inerzia definibile solo mediante l'energia e la quantità di moto acquisita dalla sorgente. Infatti, per i principi di conservazione dell'energia e della quantità di moto, durante la fase *alfa* l'interazione converte tutta l'energia e la quantità di moto trasportate dalla coppia lungo l'asse del dipolo nelle corrispondenti grandezze elettromagnetiche. Le particelle perdono quindi la loro identità inerziale, acquistando le caratteristiche ondulatorie di un onda di dipolo. Ammettendo che le particelle abbiano entrambe la stessa massa inerziale $m$, la conversione assegna ad ognuna delle particelle in collisione un ruolo simmetrico di osservatore e particella incidente la cui energia totale $mc^2$ è convertita in una sorgente di lunghezza d'onda:

$$\lambda = \frac{h'}{mv} \frac{2\beta(1-\beta\cos\theta)}{1-\beta^2}, \qquad (4)$$

in accordo con entrambi i comportamenti corpuscolare (relativistico) e ondulatorio (quantistico) della materia. Si può notare come per angoli di incidenza prossimi a zero (fig. 2) e per velocità relativistiche ($\beta \approx 1$), la lunghezza d'onda della sorgente prodotta, misurata da una qualunque delle particelle, converge alla lunghezza di de Broglie della particella compagna di massa $m$.

Dato che una coppia di cariche è per definizione formata da una carica e dalla sua anticarica, dal momento in cui una coppia è creata, ciascuna carica acquista spin (cfr. par.8), mantenendo un legame stabile e perpetuo con la sua compagna. Se il legame venisse a mancare, la sorgente cesserebbe di esistere ed entrambe le particelle non sarebbero più percepibili. Questo comporta che in un Universo neutro formato da coppie, trascorso un tempo adeguato, ogni particella entra in collisione con ogni altra anti-particella di caratteristiche identiche a quelle della propria compagna, mettendo in comune con questa solo l'energia e la quantità di moto che caratterizzano la formazione della sorgente nella direzione di collisione. La formazione di più sorgenti, converte le caratteristiche inerziali di ciascuna particella in una sovrapposizione di onde ciascuna di lunghezza d'onda (4), che globalmente descrivono la particella materiale come un pacchetto d'onde [7-8].

### 7. - Il principio d'indeterminazione

Il fatto che le sorgenti possano avere lunghezze d'onda qualunque senza perdere le loro caratteristiche, implica che un osservatore macroscopico possa essere incluso nel volume interno al fronte d'onda della sorgente.

Prendiamo in considerazione una sorgente con lunghezza d'onda nell'intervallo delle onde radio ed energia locale $E = \hbar c / \lambda$. Un osservatore interno al fronte d'onda, tale da avere una distanza di osservazione $r \leq \lambda$ misurata dal centro virtuale della sorgente, può misurare un'energia e un momento del campo elettromagnetico necessariamente sempre minori o uguali a quelli localizzati durante la formazione della sorgente all'interno del primo fronte d'onda sferico. Per l'osservatore, il prodotto tra momento localizzato e distanza propria dal centro della sorgente definisce un principio d'indeterminazione per osservatori interni:

$$P\,r \leq \hbar. \qquad (5a)$$

Viceversa, se l'osservatore non può essere contenuto nella sorgente perché la lunghezza d'onda è troppo piccola per contenerlo, questo accade per lunghezze d'onda minori o uguali a quelle della regione sub-radio, l'osservatore percepirà la sorgente come un quantità elementare, al disotto della quale non è possibile in alcun modo scendere. Questo è quanto accade per ogni possibile osservazione effettuata nel mondo delle interazioni elettromagnetiche microscopiche; ogni sistema di misurazione esterno alle sorgenti non potrà che misurare azioni maggiori uguali all'azione fondamentale di una sorgente. Per tali osservatori vale il principio d'indeterminazione

$$P\,r \geq \hbar, \qquad (5b)$$

in accordo con quello di Heisenberg [2,5].

Le indeterminazioni (5a-b), dimostrano come la quantizzazione sia un fenomeno prodotto dal differente ordine di grandezza del sistema osservato (microscopico) rispetto a quelle del sistema di osservazione (macroscopico) e non da principi fondamentali basati su differenze fenomenologiche a priori tra la fisica del microscopico e del macroscopico.

### 8. – Spin ed effetti superluminali

La componente trasversale del vettore di Poynting, descrive la propagazione di un'onda in rotazione all'interno della zona sorgente. Considerando le simmetrie del campo elettromagnetico, il momento angolare associato alla sorgente risulta dipendente dall'asse di simmetria rispetto al quale si esegue la misurazione dello spin, quindi dipendente dal sistema di osservazione.

Il campo associato alla componente trasversale del vettore di Poynting è formato da due lobi simmetrici opposti, due osservatori che eseguono la stessa misura su lati speculari della sorgente misurano quindi sempre spin opposti, in quanto il momento angolare di ciascun lobo non è invariante per lo scambio delle cariche del dipolo. In unità d'azione, lo spin per ogni osservatore è in accordo con quanto previsto dalla Meccanica Quantistica per due elettroni: ± 1/2. Mantenendo lo stesso asse di simmetria, ponendoci al centro della sorgente ed estendendo il calcolo del momento angolare all'intero campo, le componenti di spin dei due lobi si sommano dando spin zero indipendentemente dallo scambio di posizione delle cariche del dipolo.

Consideriamo ora l'emissione elettromagnetica della sorgente osservata nella direzione di propagazione dell'onda. L'asse di emissione coincide con l'asse del vettore **k** normale all'asse di dipolo. Per eseguire la misurazione del momento angolare di un onda questo è l'asse naturale per un osservatore esterno. Anche in questo caso esistono due possibili risultati dovuti alla direzione di emissione, quindi di osservazione su lati speculari della sorgente. La misurazione dello spin in base alla direzione di osservazione fornisce come risultato ± 1, corrispondente alle due differenti polarizzazioni dell'onda emessa.

Tutti i risultati ottenuti sono spiegabili in un unico quadro generale. Se consideriamo che il momento angolare associato a ciascuno dei due lobi della sorgente è in realtà associato a carica e anticarica del dipolo. Nel caso di una coppia di elettroni si ha che ciascun elettrone / anti-elettrone potrà avere spin +1/2 o -1/2 ma mai uguali valori. Combinando i valori ottenuti per lo spin di ciascun elettrone si ottiene spin zero che corrisponde per la sorgente ad uno stato bosonico che media l'interazione diretta tra gli elettroni; di fatto è il fotone virtuale scambiato nell'interazione di una coppia di elettroni. Viceversa, nel caso di osservazione dell'onda emessa, il dipolo si trova in fase *omega*, quindi l'energia e la quantità di moto delle particelle vengono trasportate sotto forma di onde con spin +1 o -1 che polarizzando il mezzo circostante danno origine a fotoni reali secondari.

Queste ultime considerazioni ci riportano al problema della comunicazione a velocità infinita tra due particelle. Dato che è possibile la creazione di una sorgente a qualunque distanza d'interazione senza che il valore della costante d'azione si modifichi, un'interazione che comporti un cambiamento dello spin di un elettrone da +1/2 a -1/2 può avere in linea di principio l'unico effetto di produrre un cambiamento causale ed istantaneo nello spin dell'elettrone compagno indipendentemente dalla distanza a cui le cariche si trovano. Diversamente, si produrrebbe un disaccoppiamento istantaneo delle particelle che potrebbe verificarsi solo violando tutti i principi di conservazione. Proprio questo risultato potrebbe essere alla base della comprensione dei fenomeni connessi alla disuguaglianza di Bell.

## 9. – Effetti cosmologici in Bridge Theory: materia oscura e radiazione cosmica di fondo

L'annuncio dell'osservazione indiretta della distribuzione 3D della materia oscura, utilizzando l'effetto di lente gravitazionale prodotto da concentrazioni non visibili di materia, permette di fare alcune considerazioni sulle potenzialità della Bridge Theory nel descrivere in un quadro teorico unico e consistente, la possibile natura microscopica e macroscopica della materia oscura. Inoltre la capacità di far coesistere contesti teorici differenti come Meccanica Quantistica e Relatività, tutti riconducibili in Bridge Theory ad un unico quadro fenomenologico, facilita sicuramente il compito di accordare formalismi solo apparentemente incompatibili.

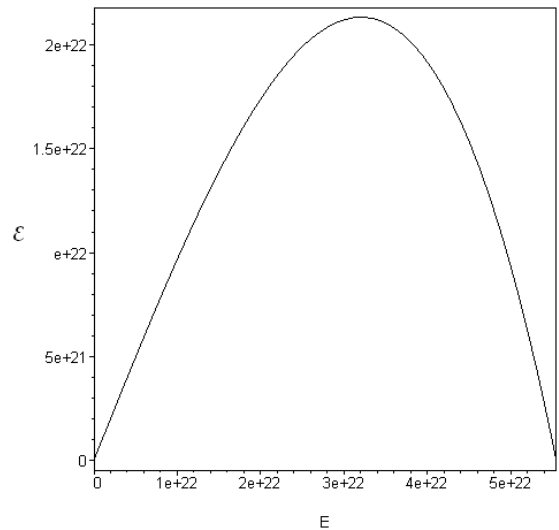

Figura 5

Per quanto detto nel cap. 6, la presenza nell'Universo di uguali quantità di carica positiva e negativa produce sorgenti dipolari in uno spazio che per quanto grande possa essere è comunque limitato dal tempo trascorso dal tempo zero ad oggi. Le sorgenti seguendo la statistica di Bose-Einstein si comportano come un gas di fotoni. Lo spettro che se ne ottiene, a differenza di quello previsto per il corpo nero, è tagliato superiormente e inferiormente a energie corrispondenti al minimo e al massimo valore consentito della lunghezza d'onda, quindi della minima distanza d'interazione tra le cariche della sorgente. Il limite energetico inferiore dello spettro $\mathcal{E}_{inf} = 2\pi\hbar/\lambda_{max}$ è in relazione con il tempo necessario ad una carica per entrare in collisione con la compagna posta alla massima distanza compatibile con il raggio dell'Universo. Il limite superiore $\mathcal{E}_{sup} = 2\pi\hbar/\lambda_{min}$ è invece in relazione con la massima energia raggiungibile da una sorgente prima che il campo gravitazionale prenda il sopravvento tagliando superiormente l'estensione dello spettro (Fig. 5).

Ulteriori incrementi dell'energia nel centro di massa della sorgente, non giustificabili nell'Universo attuale, produrrebbero un red-shift tale da far collassare progressivamente la sorgente sino a scomparire sotto l'orizzonte degli eventi di Swartzchild. Questo accadrebbe per energie della sorgente corrispondenti a lunghezze d'onda minori o uguali alla lunghezza di Planck, [10-11]:

$$\mathcal{E} = \left(1 - \Gamma \frac{E^2}{E_P^2}\right) E . \qquad (6)$$

L'energia emessa (6) evidenzia tre zone caratteristiche: una lineare che si estende sino ad energie nel centro di massa di $2.03 \ 10^{22}$ Mev, qui il red-shift è trascurabile e l'energia emessa è praticamente uguale a quella localizzata, la sorgente è in grado di emettere luce e materia; una critica che si estende dal limite della zona di linearità sino a $3.20 \ 10^{22}$ Mev: a questo valore corrisponde il limite superiore dell'energia disponibile in un'unica emissione, $2.14 \ 10^{22}$ Mev è il top della scala assoluta delle energie in presenza di effetti auto-gravitazionali della sorgente; segue una zona di collasso nella quale non si hanno ulteriori incrementi dell'energia disponibile e per $5.55 \ 10^{22}$ Mev la zona sorgente precipita sotto il proprio orizzonte degli eventi. La sorgente si trasforma in un Micro-Black-Hole (MBH) rilevabile solo gravitazionalmente.

Le caratteristiche di un MBH sono quelle di un bosone neutro di massa $9.89 \ 10^{-8}$ Kg spin zero e diametro pari alla massima estensione della sua zona sorgente $1.67 \ 10^{-35}$ m .

Ipotizziamo che queste singolarità siano le componenti principali della materia oscura, dovrebbero allora essere state prodotte in un periodo anteriore al primo minuto di evoluzione del nostro Universo. Infatti solo allora l'alta temperatura può aver favorito sia la formazione che l'instabilità dei MBH, che per le piccole dimensioni (lunghezza di Planck), possono aver iniziato una rapida evaporazione mediante l'emissione di lampi gamma e materia. Le evidenze più remote potrebbero essere proprio le micro-onde del fondo cosmico, di cui i Gamma-Ray-Burst costituiscono le evidenze più attuali. Il raffreddamento progressivo a temperature prossime allo zero assoluto, può quindi aver ridotto la frequenza di evaporazione, favorendone l'agglomerazione per via gravitazionale in condensati di Bose-Einstein, solo indirettamente osservabili tramite i loro effetti gravitazionali e le emissioni gamma prodotte durante la rapida evaporazione di parti dell'agglomerato.

La materia oscura di fatto è la trama del nostro Universo nel quale la materia ordinaria è un sottoprodotto ultimo e marginale.